# Dark Patterns and Consumer Protection Law for App Makers

Gregory M. Dickinson
*College of Law*
*University of Nebraska*
Lincoln, NE, USA
gdickinson3@unl.edu
ORCiD: 0009-0006-8992-3929

*Abstract*—Dark patterns in online commerce, especially deceptive user interface designs for apps and websites, undermine consumer autonomy and distort online markets. Although sometimes deception is intentional, the complex app development process can also unintentionally produce manipulative user interfaces. This paper discusses common design pitfalls and proposes strategies for app makers to avoid infringing user autonomy or incurring legal liability under emerging principles of consumer protection law. By focusing on choice architecture and transparent design principles, developers can both facilitate compliance and build user trust and loyalty.

## I. Introduction

The digital realm, characterized by a rapid evolution of e-commerce and marketing strategies, has fundamentally reshaped how consumers interact with sellers of goods and services [1], [2]. Artificial intelligence and machine learning underpin much of this transformation, enabling product personalization and targeted marketing at a large scale [3], [4], [5].

This technological advancement, however, while offering immense benefits, has also paved the way for the pervasive growth of online fraud, sometimes accomplished via so-called "dark patterns." Dark patterns are user interface design structures that benefit their creators, typically app and website owners, by nudging, steering, or outright deceiving users into making decisions preferred by the UI designer, sometimes contrary to users' own intent or better judgment [6], [7]. They represent a critical challenge to the integrity and trustworthiness of the digital realm, as they substitute trickery and deception for the user autonomy and free choice that should drive the market for digital products [8], [9].

The proliferation of dark patterns is thus a major concern for consumers. Empirical studies reveal their widespread presence, but results must be interpreted cautiously given the imprecision of the term dark patterns, which can describe techniques ranging from mere user inconvenience or annoyance to outright fraud. For instance, a 2020 study by Geronimo et al. found that 95% of 240 popular mobile applications contained one or more forms of dark patterns, with an average of seven different deceiving interfaces per app, but considered an app to include a dark pattern whenever its "[user interface] and interaction seem to benefit the system instead of the user" [10].

That definition makes some sense for paid products, but is much less appropriate for evaluating free software products, which depend on a symbiotic advertising and data-sharing arrangement between app maker and user to deliver useful products at no out-of-pocket cost [46], [47], [48], [49]. Studies such as Geronimo et al. thus shed light on design trends, but cannot measure the prevalence of *unlawful* designs, for their definition of dark patterns is far more inclusive than the legal standards for fraud and deceptive consumer practices, which require intent to deceive or the likelihood of deceiving (not merely annoying or inconveniencing) a reasonable consumer.

Many website and app makers, however, will want to avoid dark patterns of all varieties. Although many users do not initially recognize these manipulative designs, awareness significantly increases once they are informed about the issue [10], leading to a general dislike and mistrust of services employing such tactics [12], [13]. Moreover, the use of aggressive, deceptive dark patterns has generated a powerful backlash among consumers by eroding user trust in digital products [13], [14].

The legal and regulatory landscape is changing in response to this phenomenon. Governments and consumer protection agencies, particularly in the United States, are taking a more proactive stance against deceptive design practices. The Federal Trade Commission (FTC) and state attorneys general are more frequently initiating investigations and enforcement actions against companies whose apps and websites employ dark patterns [15], [16]. Such enforcement actions follow naturally from the premise that if a user interface or website is likely to mislead a reasonable user, it is unlawful under the FTC Act and state consumer-protection laws, which provide that deceptive designs can be unlawful, even if unintentional [17], if they are likely to mislead a reasonable consumer [18]. This rising legal scrutiny underscores the need for app makers to understand and mitigate the risks associated with dark patterns, whether intentionally or accidentally created.

## II. Background

The concept of dark patterns originated in the broader discourse of design patterns and anti-patterns within software engineering and user interface (UI) user experience (UX) design theory. Understanding this evolution provides crucial context for appreciating the modern problem of dark patterns.

### A. From Design Patterns to Dark Patterns

The notion of "patterns" emerged initially in architecture as a means of capturing and sharing proven solutions to recurring problems. Examples include terraces to provide private spaces in public areas and playgrounds to give children suitable spaces for play [19]. In user interface design, this idea took hold in the 1990s, with the aim of documenting reusable, effective design elements that enhance usability and learnability [20]. Examples include common UI components like checkboxes, radio buttons, and navigation menus that users instinctively understand and interact with, facilitating intuitive digital experiences [20]. The goal of a design pattern is to provide a blueprint for good design, leading to flexible and maintainable systems [21].

Soon after the rise of design patterns in software development, the related concept of "antipatterns" also came to prominence, in the late 1990s, particularly within the context of object-oriented programming [22]. Antipatterns identify commonly used but ultimately suboptimal solutions to recurring problems [22]. These are design choices that, while perhaps offering a quick fix, lead to systems that are difficult to maintain, build upon, or scale over time [23]. They are often the result of insufficient knowledge or experience, or of the misapplication of a good design pattern in the wrong context [23], [24]. Because success in software development is often determined by the *absence* of problems, antipatterns are useful in helping developers to identify and avoid common missteps [24], [25].

Building upon the antipattern idea, the concept of dark patterns emerged to describe a particular subset of poor design choices [6], [26]. Unlike generic antipatterns, dark patterns are not merely inconvenient or unscalable or poor programming practices. Instead, they are *designed* to trick, pressure, or inconvenience the user to achieve the company's goals, often at the expense of the user's best interests [6], [27]. As Mirnig and Tscheligi elaborate, dark patterns differ from antipatterns primarily in that they are designed intentionally to operate contrary to the user's interests [28]. This shift marks a critical theoretical recognition that UI design can be deliberately used not just to guide, but to mislead.

### B. Manifestations and Academic Discourse

Dark patterns manifest in various forms. Like traditional, in-person sales techniques they often leverage information asymmetries, user search costs, and deceptive phrasing to influence user behavior [13], [29], [30]. These tactics can be subtle, making them difficult for users to detect, or they can be aggressive, prompting a strong negative reaction [13]. Researchers have categorized these patterns to better understand and combat them. Brignull's initial work identified a set of common techniques, which has since been expanded and refined by others. [31].

Brignull offered a twelve-pattern typology in the early 2010s that distinguishes patterns by the contexts in which they appear and the ways they deceive users. [31] Christoph Bösch and others later built on Brignull's work to offer a seven-pattern typology along with examples and potential countermeasures [27], and, most recently, Colin Gray and his coauthors condensed Brignull's categories into a compressed, five-pattern typology (depicted below) organized around patterns' modes of deception: Nagging, Obstruction, Sneaking, Interface Interference, and Forced Action [32]. To understand the phenomenon of dark patterns, however, it is helpful to consider just a few of many more particular examples.

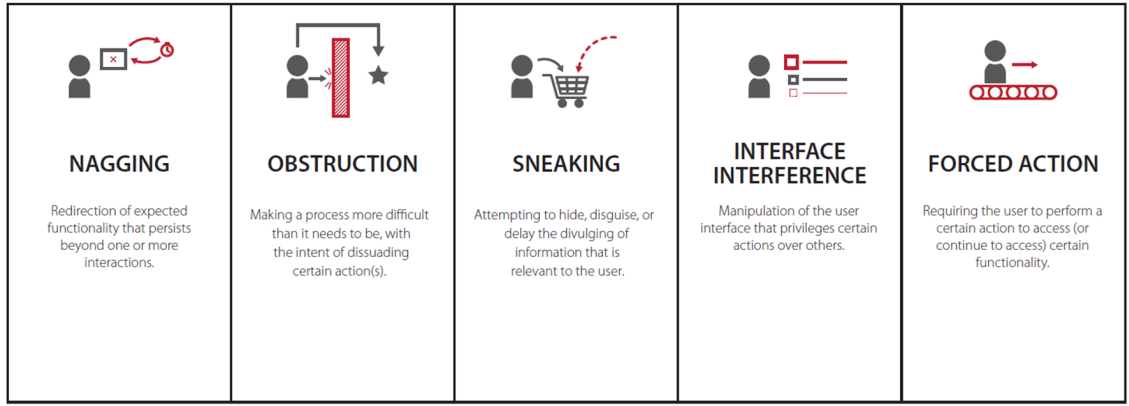

Fig. 1. Dark-Pattern Typology from Gray et al. (2018)

Nagging-type dark patterns involve repeated, persistent requests to the user, often for optional actions that the app maker desires. For instance, many mobile applications frequently display pop-ups asking users to rate the app in the app store, enable push notifications, or connect their social media accounts. While these actions are typically optional, the incessant nature of these prompts, sometimes reappearing even after a user has dismissed them multiple times, can create the impression that the action is obligatory rather than optional. This unexpected functionality (optional, but persistent) that extends across several interactions is a key characteristic of nagging [7].

Obstruction-type dark patterns are those that make it unnecessarily difficult for users to accomplish the objectives they seek. A common example, sometimes called "roach motels" are UIs that make it easy for users to get into a situation (e.g., subscribing to a service or initiating a free trial) but difficult to exit [7]. Often that involves hiding cancellation options, requiring multiple clicks or phone calls, or otherwise obscuring the steps required to unsubscribe. Amazon Prime's multistep cancellation process, which has faced legal challenge by the FTC, is an especially prominent example of this type of obstruction [33].

Sneaking-type dark patterns are those that hide or delay divulging information that is relevant to a user's decision. The most common examples are hidden costs, where unexpected fees (e.g., delivery charges, taxes, service fees) are revealed late in the purchase process, often just before final payment [8]. By the time these costs appear, users have already invested significant time and effort in the transaction (e.g., creating a user account or selecting a particular model), making it more time efficient for them to complete the transaction than to begin the process again with a new seller, even though they would not have chosen the seller if complete information had been available at the outset.

Two prevalent techniques falling within the Interface-Interference type are deceptive countdown timers and false scarcity [7]. These types of patterns create an artificial sense of urgency or limited availability, pressuring users into making quick decisions based on information that is both limited (no time for further research) and false (not truly an urgent decision or scarce resource). These techniques favor the app maker at the expense of users by undermining consumers' efforts to make welfare-maximizing purchasing decisions. Mathur et al. observed such deceptive countdown timers on 140 shopping websites, where the advertised offer remained valid even after the timer expired, suggesting an intent to mislead [8]. Similarly, "limited stock" notifications, when untrue, drive transactions by creating in users a false fear of missing out on a bargain [13].

Another Interface-Interference-type pattern is a technique sometimes known as "confirmshaming," which attempts to invoke guilt in the user, hoping she will take a particular action in response. [13]. This involves phrasing one option in a way that will make the user feel bad or inferior for selecting it, such as "No thanks, I don't want to save money" or "I'm not interested in helping others." Yet another Interface-Interference-type pattern involves designing interfaces so that users are likely to make unintentional transactions. A leading example is Epic Games, creator of the popular videogame Fortnight, which in 2022, settled an action initiated by the FTC alleging that deceptive user interface features of Fortnight had led users, many times children, to make unintended purchases with the click of a single button [34].

Finally, Forced-Action patterns are any UI design that forces, or strongly encourages, users into taking actions preferred by the app maker, such as permissive sharing settings. Often this is accomplished through default data-sharing settings or by making privacy controls complex and hard to navigate. One notable instance, cited by Narayanan et al., involved Facebook asking users for phone numbers for two-factor authentication, only to then use those numbers for targeted advertising [6]. Consent banners are another common context for Forced-Action dark patterns, where designers exploit users' time scarcity and consequent reluctance to engage with complex privacy settings to pressure them into permissive data-sharing settings [7].

Prior academic work on dark patterns includes developing taxonomies to name dark patterns and empirical studies to measure their prevalence and effectiveness. Research by Mathur et al. conducted a large-scale crawl of 11,000 shopping websites, identifying 1,818 dark pattern instances across 15 types and 7 broader categories. Their findings revealed that 183 websites engaged in deceptive practices, and alarmingly, 22 third-party entities offered dark patterns as "turnkey solutions" for businesses [8]. This highlights the commercialization and systemic integration of deceptive designs.

Luguri and Strahilevitz's empirical studies provide compelling evidence of the power of dark patterns, showing that even mild dark patterns can significantly increase user engagement with dubious services, particularly among less educated subjects [13]. Their work identified hidden information, trick questions, and obstruction strategies as particularly effective in manipulating consumers [13].

Some academic work also extends to the ethical implications of dark patterns. Gray et al. [7] emphasize that UI designers can easily become complicit in manipulative practices, underscoring the need for a broader research focus on the ethics of user experience. Mathur, Mayer, and Kshirsagar [28] call for a more conceptual foundation for analyzing dark patterns, drawing on insights from psychology, economics, ethics, philosophy, and law to articulate normative perspectives and empirical measurement methods. They argue that future research should move beyond subjective criticism to apply methods grounded in these long-standing normative frameworks.

### C. The Legal Landscape

From a legal perspective, dark patterns are most appropriately scrutinized under the doctrines of consumer protection law, particularly common-law fraud and state and federal statutory prohibitions on deceptive business practices. Under modern law, the basic legal principle, under both the Federal

Trade Commission Act (FTC Act) and similar state statutes, is that an entity is prohibited from engaging in "deceptive acts or practices in or affecting commerce" [33]. For present purposes, the most important feature of this legal standard is that intent to deceive is *not* required for conduct to be deceptive and unlawful. [18]. A UI design is unlawful if it is *likely to mislead* a reasonable consumer, even if the misleading outcome was accidental [35], [18]. This "reasonable consumer" standard focuses on the *effect* of the design, which may differ from the designer's subjective intent [35].

This legal framework is particularly important given the sophisticated nature of modern digital marketing. As discussed by Calo and Dickinson, digital marketing, driven by personalization and data analytics, introduces new complexities that traditional legal frameworks must address [16], [36]. The widespread collection of consumer data, the customization of digital interfaces, and the use of machine learning and AI in design processes (e.g., A/B testing) amplify the potential for deception [37], [3], [5]. Willis goes so far as to argue that for the law to demand evidence even of *how* a consumer was deceived or which materials did the deceiving would place an "impossible burden of proof" on enforcement agencies when dealing with "machine-created, micro-targeted digital deception" [38]. Deceptive *outcomes* and the likelihood (ex ante) of a user being deceived, regardless of intent, are the concerns of modern consumer law and therefore the key standard for developers to bear in mind.

The FTC has pursued cases against companies that employ deceptive dark patterns. Beyond the Amazon Prime example, the FTC has also combatted UIs that press users into automatic subscription renewals that are difficult to cancel [39]; and, in another prominent dispute, the FTC argued that Intuit's deemphasis of its free TurboTax product in favor of paid versions might constitute a deceptive trade practice [40]. State Attorneys General have encountered similar tactics at the state level, chronicling problematic behaviors in recent correspondence with the FTC, and emphasizing how existing standards bar deception regardless of the technological tools employed [15].

Although the legal framework may evolve to become more detailed regarding specific practices, the core principles have remained the same over time: Like other modes of communication, UI designs that lead a reasonable consumer to make unintended or misinformed choices are subject to legal challenge. The difficulty for app makers, however, is not merely to refrain from intentionally employing dark patterns, but also to recognize and rectify those deceptive UIs that arise inadvertently during the complex software-development process.

### III. ACCIDENTAL DARK PATTERNS

While many discussions of dark patterns emphasize their malicious, intentional creation, it is crucial for app developers and companies to recognize that problematic UI designs can also arise *accidentally* during the app development process. The complexity of modern software development and the pressures of optimizing for certain metrics can lead to designs that, while not intended to deceive, are nonetheless misleading to reasonable consumers thus unlawful pursuing to state and federal consumer protection provisions [17].

#### *A. The Unintended Consequences of Design*

The iterative nature of app development, often involving rapid prototyping and A/B testing, creates an environment where dark patterns can emerge without any ill intent toward users. Designers might, for example, optimize for conversion rates or engagement metrics without fully scrutinizing the clarity of the resulting user experience. A design that performs "best" in A/B tests in optimizing a relevant metric might do so because it confuses users rather than because it improves their experience or the app's choice architecture [41]. As Gray et al. note, UI designers face increasing legal scrutiny, and without careful ethical consideration, they can become "complicit in manipulative or unreasonably persuasive practices" [7]. The legal standard, as discussed above, focuses on the *likelihood of misleading* a reasonable consumer, not the intent of the designer, making accidental dark patterns a significant risk.

#### *B. Common Dark Pattern Pitfalls*

Below are examples of dark patterns that are easy to create inadvertently and (rather than being merely annoying) are also misleading enough potentially to violate legal prohibitions on deceptive commercial practices.

*1. Confusing Wording, Double Negatives, and Unclear Discount Terms*

*Description*: This type of accidental dark pattern arises from poorly crafted UI text, complex legalistic language, or attempts to be clever with phrasing that results in unclear meaning. Double negatives (e.g., "Do you *not* wish to opt out?") are classic examples that can easily confuse users into making choices that are the opposite of their intention. Similarly, sales terms or eligibility conditions for discounts that are verbose, hard to find, or rely on convoluted logic can unintentionally mislead consumers about the true nature or value of an offer.

*Mechanism of Accidental Creation*: Designers, often under pressure to condense information or to craft attractive phrasing, might inadvertently introduce overly complex language into a UI. Similarly, A/B testing, if focused solely on conversion rates (e.g., how many users click "yes"), without deeper qualitative analysis of *why* users are clicking, can incentivize unclear phrasing. In an extreme case, for example, a test might show that swapping "yes" and "no" buttons for marketing list enrollment increases opt-ins, but the increase would be due to consumer confusion, not genuine desire [10]. If a designer does not understand why a UI design performs well in A/B testing, she might implement a "successful," but deceptive design [36]. The result is that users might sign up for services or agree to terms that they do not properly understand or believe that they qualify for discounts that they do not.

*Legal Implications*: Confusing wording directly impacts the "reasonable consumer" standard, for if a significant number of users is likely to misinterpret its wording, then a UI design could be deemed deceptive under the FTC Act and analogue state consumer protection statutes. [15], [35], [42].

*2. Overly Frequent Reminder Prompts for Optional Actions*

*Description*: Many apps use prompts to encourage optional user actions such as enabling notifications, rating the app, linking social media accounts, or leaving a tip for a service provider. Although occasional, well-timed prompt are tolerable and sometimes helpful, excessive or persistent reminders can cross the line into "nagging," a recognized dark pattern [32] that may rise to the level of a deceptive and therefore unlawful design if a reasonable user might inaccurately conclude that the prompted action is not optional.

*Mechanism of Accidental Creation*: Developers and product managers might implement frequent prompts based on internal metrics (e.g., increasing notification opt-in rates) without fully considering the user's perception. The intent might be to simply "remind" users of a beneficial feature. However, continuous prompting for an *optional* action can lead users to infer that an action is, in fact, mandatory for continued use of the app or to avoid negative consequences. For example, a user repeatedly prompted to enable location services might infer (perhaps inaccurately) that the app will not function correctly without location services enabled. Kollmer and Eckhardt provide an example of Instagram's persistent dialogue for activating app notifications, where the only options provided are "Not Now" and "OK" [9]. The design is a borderline case. Its language presents the notification setting as optional (as it truly is), but the app's persistence, even after "Not Now" has once been selected, and the lack of a "Never" option could lead a user to conclude that consent is required.

*Legal Implications*: When a UI design presents an optional action to its users in a way that could lead a reasonable user to conclude that the action is required, it violates state and federal legal prohibitions against deceptive trade practices. Such UI designs cross the line between persuasion and compulsion by extracting from the user a "voluntary" action while she misapprehends the relevant facts.

*3. Misleading or Outdated Claims Regarding Privacy or Security*

*Description*: This category includes instances where an app's interface or accompanying descriptions make claims about its features, security measures, or privacy protections, or other characteristics that are no longer accurate, incomplete, or open to misinterpretation. Examples include vague statements about "encrypted data" or "state-of-the-art credit card transmission security" that users might interpret as blanket guarantees of robust security practices across the board, even if other vulnerabilities exist or if the described feature applies only to a specific part of the system. Wiener's work on "deception that notifies greater privacy protection than is actual implemented" is instructive here, where designs seem to offer user protection but actually impede it [16].

*Mechanism of Accidental Creation*: As apps evolve, features and underlying technologies change. Marketing copy and UI text, however, may not be updated promptly. An old security claim might linger in the app's description even after the app's approach to security changes. Knowledgeable developers might use generic, reassuring, or highly technical terms without realizing that those terms might be misinterpreted by users as absolute guarantees.

*Legal Implications*: The FTC has a strong history of enforcement actions against companies making misleading security and privacy claims. For instance, in a notable case, the FTC charged a mobile app developer for misrepresenting that its flashlight app would access photos and location data only when in use, while in reality, it transmitted users' precise location data to third-party ad networks whenever the device was powered on [43]. This and similar practices are obvious instances of a deceptive practice under Section 5 of the FTC Act given the discrepancy between stated policy and actual behavior.

4. Placing Buttons in Hard-to-Find Locations for Disfavored Actions

*Description*: This dark pattern involves burying critical user functionality—such as canceling a subscription, deleting an account, or finding a free version of a service—deep within menus, requiring an excessive number of clicks or navigating of obscure pathways. The intent may not be malicious, but rather a decision driven by business goals to reduce cancellations or encourage paid subscriptions.

*Mechanism of Accidental Creation*: Business objectives often influence design decisions. Teams might be incentivized to reduce cancellations or increase premium sign-ups. Without strict oversight or incentive mechanisms designed to ensure developers' incentives remain aligned with the interest of the entity, designers might make disfavored actions less prominent or more cumbersome, maximizing their team's objective while nonetheless degrading the overall user experience. Shortsighted leadership often leads to Obstruction-type dark patterns, where the difficulty of finding an option discourages its use [13]. A design might not explicitly hide the option, but its placement and required effort effectively make it difficult for a reasonable user to discover. Examples like Amazon Prime's multi-step cancellation process, which was the subject of FTC scrutiny, and TurboTax's alleged hiding of its free tax-filing program illustrate how locally optimized design decisions can impede the overall user experience and even create legal liability for deceptive practices [8], [18].

*Legal Implications*: These UI designs constitute "deceptive acts or practices" if they create an unreasonable obstacle to consumer choice (e.g., to cancel a service or access a free alternative). The FTC and state regulators view such UI designs—easy to get in, hard to get out—as deceptive because they undermine a consumer's ability to easily reverse a decision or exercise a choice she might have made if the available choices had been readily apparent [15] [35] [42]. The focus is on whether the design makes it "difficult for users to express their actual preferences" [12].

5. Accidental Transactions

*Description*: This dark pattern makes it too easy for users to unintentionally complete transactions, such as adding unwanted items to their shopping carts. This can happen because of preselected checkboxes for additional items or services, automatic addition of "recommended" items to a cart without explicit user consent, or overly sensitive "buy now" buttons that are prone to accidental taps.

*Mechanism of Accidental Creation*: The desire to create a seamless user experience and reduce friction in the purchase process can lead developers to inadvertently create UIs that employ this dark pattern. Developers might, for example, implement default options or "smart" features (like auto-adding related items) with the intent of being helpful or accelerating the checkout process. However, if these features are not transparently communicated or do not require clear, explicit user action, they become deceptive. Dark patterns of this sort can result in hidden costs or unwanted purchases, as users might overlook the automatically added items before completing their transaction [32], [8].

*Legal Implications*: Practices that lead to accidental purchases or subscriptions are a clear target for the FTC and state consumer protection enforcement agencies. Indeed, the FTC has promulgated special rules governing "negative option" plans where goods or services are automatically provided unless the consumer explicitly opts out [44]. A UI design that makes it too easy to inadvertently opt-in or to add unwanted items to a purchase is likely to mislead reasonable consumers and therefore runs afoul of consumer protection law [13].

These five examples illustrate that dark patterns are not always the result of malevolent intent. Often, they are the byproduct of shortsighted UI optimization strategies or poor design oversight. Regardless of intent, however, the legal ramifications can be significant.

IV. BEST PRACTICES TO AVOID DARK PATTERNS

Given that dark patterns can be created entirely by accident, it is imperative that companies train their developers to be alert for inadvertent dark patterns. Relying solely on the absence of malicious intent is insufficient, as consumer protection law focuses on the *effect* of the design on a reasonable consumer. By implementing design principles that align the incentives of app developers with broader consumer-satisfaction goals, companies can significantly reduce their risk of inadvertently creating unlawful dark patterns, while simultaneously fostering user trust and loyalty.

A. General Principles: Cultivating a User-Focused Design Mindset

The first and most critical step is to foster a user-focused mindset throughout the app development cycle. This involves:

*Prioritizing User Autonomy and Control*: UI design should empower users to make informed and free choices, not manipulating them into taking a particular path. Genuine respect for user freedom of choice will attract those users truly interested in the product or service on offer and increase user satisfaction and retention [9], [16], [45].

*Transparency by Default*: Information, particularly concerning costs, terms, and data collection should be presented clearly, concisely, and up front. As Sandhaus highlights, "bright patterns" prioritize transparency and explainability, countering the obfuscation often found in dark patterns [37].

*Education and Training*: Programmers, UX/UI designers, product managers, and even marketing teams need explicit training on what constitutes a dark pattern, how they can arise unintentionally, and their potential legal consequences [40]. This training should go beyond theoretical concepts to practical examples and case studies relevant to their work.

*Internal Design Audits*: Companies should establish regular, systematic reviews of their interfaces specifically to identify potential dark patterns, both intentional and accidental. These audits should involve diverse perspectives, including legal counsel and user advocates.

*Shift from "Conversion at Any Cost"*: Instead of optimizing for clicks or sign-ups, teams should also track user satisfaction, retention rates, which are driven by genuine value, and explicit user feedback regarding clarity and control. Such controls not only help companies comply with legal requirements, but will also produce greater consumer satisfaction, reducing the risk of aggressive new lawmaking, which could harm both consumers and businesses [18].

B. Specific Best Practices Addressing Accidental Dark Pattern Pitfalls

Applying the general principles above to the five accidental dark pattern categories discussed in Section III yields several concrete best practices for programmers and designers:

1. Avoiding Confusing Wording

*Plain Language and Clarity*: All UI text, prompts, and legal disclaimers should be written in clear, unambiguous language that is easily understood by the target audience. Avoid jargon, overly complex sentences, and especially double negatives.

*User-Centric Phrasing*: Focus on what the *user* wants to accomplish, not the company's preferred outcome. For example, instead of "Do not decline if you wish to receive notifications," use "Receive Notifications: Yes / No."

*Transparency in Offers*: Clearly state the conditions, eligibility requirements, and full cost of any offer or sale upfront. Use visual cues (e.g., distinct colors, prominent text) to highlight important terms rather than burying them in fine print.

*Qualitative A/B Testing Analysis*: When conducting A/B tests, go beyond quantitative metrics (like click-through rates). Supplement with qualitative feedback (e.g., user interviews or eye tracking) to understand *why* one design performs better on a quantitative metric. Ensure, for example, that an increased conversion is attributable to user demand not confusion.

2. Managing Reminder Prompts to Avoid Nagging

*Contextual and Timely Prompts*: Reminders for optional actions should be presented at relevant moments in a user's journey, not randomly or excessively. For instance, it would be better to prompt the user to submit a review of an app after successfully using it rather than every time the app is opened.

*Clear Opt-Out for Prompts*: Every prompt for an optional action should include a clear, easily discoverable "No Thanks" or "Later" option that is as visually prominent as the "Yes" option.

*Frequency Caps*: Implement technical limits on how often a user is prompted for the same optional action. If a user declines, respect that choice for a reasonable period.

*Explainable Actions*: If a prompt is tied to a specific benefit, clearly articulate that benefit (e.g., "Enable notifications to get real-time order updates"). This empowers the user to make an informed decision [9].

### 3. Ensuring Accuracy in Claims Regarding Privacy and Security

*Regular Content Audits*: Implement a process for regularly reviewing and updating all app descriptions, UI text, and privacy policies to ensure they accurately reflect current features, security practices, and data handling. This type of content quickly becomes outdated [26].

*Precision in Language*: Avoid vague or overly optimistic language about security or privacy. Be specific about what is protected and how, without implying guarantees beyond actual capabilities.

*Clear Disclaimers*: If certain features or guarantees have limitations, those limitations should be clearly and conspicuously disclosed [26].

### 4. Ensuring Easy Access to Disfavored Actions

*Discoverability and Accessibility*: Critical user actions must be easy to find and access even if the company might prefer users not to take those actions (e.g., canceling a subscription, deleting an account, finding a free tier). Such functionality should not be buried in nested menus or require an unreasonable number of clicks to access.

*Principle of Least Surprise*: Design the UI such that its behavior aligns with user expectations. If users expect a cancellation button to be in account settings, it should be there and clearly labeled.

*Streamlined Processes*: Once a user initiates a "disfavored" action, the process should be as straightforward and frictionless as possible. Avoid unnecessary steps, redundant confirmations, or forced engagements designed to deter the user.

*Prominent Free Options*: If a free version of the app or service exists, ensure it is as easy to find and access as the paid version. [8], [18]. Companies like Intuit have faced legal challenges for allegedly making their free offerings difficult to locate [40], [50], [51]. The goal should be to present all legitimate options clearly, allowing users to make unhindered choices.

### 5. Preventing Accidental Clicks and Unwanted Auto-Additions

*Explicit Consent for Purchases*: For any action that involves a financial transaction, require an explicit user action, such as a clear "Add to Cart" button or a multi-step confirmation process. Avoid prechecked boxes for additional items or services.

*Clarity Regarding Recommended Products*: If an app suggests additional items for the user to purchase, these should be presented clearly as recommendations. Additional deliberate action (e.g., a distinct "Add Recommended Item" button) should be required before the item is included in the transaction. It should be obvious to the user that the suggested item is a recommended product, not a mandatory part of the purchase.

*Confirmation Screens:* Implement confirmation screens for significant actions, especially purchases, providing a summary of items and total cost before finalization. This gives users a last chance to review and correct any accidental additions.

*Undo Mechanisms*: For actions like adding items to a cart, provide an easy "undo" or "remove" option that is readily visible and intuitive to use [43]. This empowers users to correct mistakes on their own with minimal frustration.

By proactively implementing these best practices, app makers can foster a development culture that prioritizes user-focused design. This approach not only aligns with legal requirements for consumer protection but helps to build long-term customer trust and loyalty, which are invaluable assets in the competitive digital market. User-focused design decisions lead to positive outcomes for both users and businesses [37].

## V. CONCLUSION

The proliferation of dark patterns in online applications represents a significant and growing challenge to consumer autonomy and market integrity. Deceptive user interface designs, whether intentionally crafted or inadvertently introduced through complex development processes, undermine user trust and can also incur legal liability. As the digital economy continues to expand, scrutiny by the FTC and state regulators is intensifying. Whether analogue or digital, regulators are on the lookout for all practices likely to mislead a reasonable consumer, whether or not the practice is intentional.

This paper illuminates the historical progression from general software patterns to anti-patterns and, finally, to digital dark patterns. It highlights five common categories of accidental dark patterns—confusing wording, nagging prompts, misleading security claims, hidden options, and accidental transactions—demonstrating how even well-intentioned development practices can produce unlawful UI designs. Crucially, the paper proposes a collection of best practices centered on transparency, user control, and user-focused training within development teams. By embracing these principles, app makers can not only mitigate legal risks but also foster a more trustworthy and user-centric development process, ultimately benefiting both consumers and businesses.

REFERENCES

[1] M. Faruk, M. Rahman, and S. Hasan, "How digital marketing evolved over time: A bibliometric analysis on scopus database," *Heliyon*, vol. 7, no. 12, p. e08603, Dec. 2021.

[2] A. Rosário and R. Raimundo, "Consumer Marketing Strategy and E-Commerce in the Last Decade: A Literature Review," *J. Theor. Appl. Electron. Commer. Res.*, vol. 16, no. 7, pp. 3003–3024, Jul. 2021.

[3] T. Kumar and M. Trakru, "The Colossal Impact of Artificial Intelligence in E-Commerce: Statistics and Facts," *Int. Res. J. Eng. Technol*. (IRJET), vol. 6, no. 5, pp. 69–72, May 2019.

[4] B. Kotras, "Mass personalization: Predictive marketing algorithms and the reshaping of consumer knowledge," *Big Data Soc.*, vol. 7, no. 2, pp. 1–14, Jul. 2020.

[5] L. T. Khrais, "Role of Artificial Intelligence in Shaping Consumer Demand in E-Commerce," *Future Internet*, vol. 12, no. 12, p. 226, Dec. 2020.

[6] A. Narayanan, A. Mathur, M. Chetty, and M. Kshirsagar, "Dark Patterns: Past, Present, and Future," *ACM Queue*, vol. 18, no. 2, pp. 67–91, Mar. 2020.

[7] C. M. Gray et al., "The Dark (Patterns) Side of UX Design," in *Proc. CHI Conf. Hum. Factors Comput. Syst.*, Montreal, QC, Canada, Apr. 2018, Paper 534.

[8] A. Mathur et al., "Dark Patterns at Scale: Findings from a Crawl of 11K Shopping Websites," *Proc. Priv. Enhancing Technol.*, vol. 2019, no. 4, pp. 109–131, 2019.

[9] T. Kollmer and A. Eckhardt, "Dark Patterns Conceptualization and Future Research Directions," *Bus. Inf. Syst. Eng.*, vol. 65, no. 2, pp. 201–208, Apr. 2023.

[10] L. D. Geronimo et al., "UI Dark Patterns and Where to Find Them: A Study on Mobile Applications and User Perception," in *Proc. CHI Conf. Hum. Factors Comput. Systm.*, Honolulu, HI, USA, 2020, pp. 1–13.

[12] Y. Lu et al., "From Awareness to Action: Exploring End-User Empowerment Interventions for Dark Patterns in UX," *Proc. ACM Hum.-Comput. Interact.*, vol. 8, no. CSCW1, Art. 59, Apr. 2024.

[13] J. Luguri and L. J. Strahilevitz, "Shining a Light on Dark Patterns," *J. Legal Anal.*, vol. 13, no. 1, pp. 109–176, 2021.

[14] G. M. Dickinson, "Privately Policing Dark Patterns," *Ga. Law Rev.*, vol. 57, no. 4, pp. 1633–1668, 2023.

[15] State Attorneys General, "Letter to FTC re: Dark Patterns," Aug. 2, 2022. [Online]. Available: https://www.iowaattorneygeneral.gov/media/cms/17_Attorneys_General_Hawaii_OCP_Dig_FA07C81337A62.pdf.

[16] J. Wiener, "Deceptive Design and Ongoing Consent in Privacy Law," *Ottawa L. Rev.,* vol. 53, no. 1, pp. 133–169, 2021.

[17] J. G. Hurwitz, "Designing a pattern, darkly," *N.C. J. Law Technol.*, vol. 22, no. 1, pp. 57–116, Oct. 2020.

[18] G. M. Dickinson, "The Patterns of Digital Deception," *Geo. Wash. Law Rev.*, vol. 89, no. 6, pp. 2458–2503, 2021.

[19] C. Alexander, S. Ishikawa, and M. Silverstein, *A Pattern Language: Towns, Buildings, Construction*. New York, NY, USA: Oxford Univ. Press, 1977.

[20] M. van Welie, G. C. van der Veer, and A. Eliëns, "Patterns as Tools for User Interface Design," in *Proc. PP*. 2000, pp.1-12.

[21] A. L. Correa, C. M. L. Werner, and G. Zaverucha, "Object Oriented Design Expertise Reuse: an Approach Based on Heuristics, Design Patterns and Anti-Patterns," COPPE/UFRJ - Comput. Sci. Dept., Federal Univ. Rio de Janeiro, 1998.

[22] V. Arnaoudova et al., "A New Family of Software Anti-Patterns: Linguistic Anti-Patterns," in *Proc. WCRE*, 2013, pp. 225–234.

[23] N. Doty, "Privacy Design Patterns and Anti-Patterns: Patterns Misapplied and Unintended Consequences," UC Berkeley, School of Information, 2008.

[24] C. U. Smith and L. G. Williams, "New Software Performance AntiPatterns: More Ways to Shoot Yourself in the Foot," Perform. Eng. Services, 2002.

[25] A. G. Mirnig and M. Tscheligi, "Don't Join the Dark Side: An Initial Analysis and Classification of Regular, Anti-, and Dark Patterns," in *Proc. PATTERNS 2017*, Barcelona, Spain, 2017, pp. 58–63.

[26] K. Bongard-Blanchy, et al., "I am definitely manipulated, even when I am aware of it. It's ridiculous!" — Dark patterns from the end-user perspective," in *Proc. ACM Designing Interactive Syst. Conf.* (DIS '21), Virtual Event, USA, Jun.–Jul. 2021, pp. 763–776, doi: 10.1145/3461778.3462086.

[27] C. Bösch et al., "Tales from the Dark Side: Privacy Dark Strategies and Privacy Dark Patterns," *Proc. Priv. Enhancing Technol.*, vol. 2016, no. 4, pp. 237–254, 2016.

[28] A. Mathur, J. Mayer, and M. Kshiragar, "What makes a dark pattern... dark?: Design attributes, normative consideration, and measurement methods," in *Proc. CHI Conf. Hum. Factros Comput. Syst,*, Yokohama, Japan, May 2021, pp. 1–12.

[29] C. Cara, "Dark Patterns in The Media: A Systematic Review," *Netw. Intell. Stud.*, vol. 7, no. 14, pp. 106–114, 2019.

[30] M. R. Darby and E. Karni, "Free competition and the optimal amount of fraud," *J. Law Econ.*, vol. 16, no. 1, pp. 67–88, 1973.

[31] H. Brignull, "Dark patterns: Deception vs. honesty in UI design," *Deceptive Design*, Nov. 1, 2011. [Online]. Available: https://www.deceptive.design/blog/dark-patterns-deception-vs-honesty-in-ui-design

[32] C. M. Gray et al., "Dark Patterns and the Legal Requirements of Consent Banners: An Interaction Criticism Perspective," in *Proc. DIS Conf.*, New York, NY, USA, 2020, pp. 1–15.

[33] D. Pridgen, *Consumer Protection and the Law*. St. Paul, MN, USA: Thomson Reuters, 2024.

[34] W. Grantham-Philips, "More refunds are being sent to Fortnite players 'tricked' into unwanted purchases. How you can apply," *Assoc. Press*, Jun. 26, 2025. [Online]. Available: https://apnews.com/article/fortnite-settlement-refund-ftc-sends-payments-bbe3c950414dd3a56590e6a858d2c7cd

[35] .C. Miller III, "Letter to Representative John D. Dingell, Chairman, Comm. On Energy and Commerce–FTC Policy Statement on Deception," Fed. Trade Comm'n, Oct. 14, 1983. [Online]. Available: https://www.ftc.gov/system/files/documents/public_statements/410531/831014deceptionstmt.pdf

[36] R. Calo, "Digital Market Manipulation," *Geo. Wash. Law Rev.*, vol. 82, no. 4, pp. 995–1051, 2014.

[37] H. Sandhaus, "Promoting Bright Patterns," *arXiv preprint arXiv:2304.01157*, Apr. 2023.

[38] L. E. Willis, "Deception by Design," *Harv. J. Law Technol.*, vol. 34, no. 1, pp. 115–190, 2020.

[39] Federal Trade Commission, "FTC takes action against Amazon for enrolling consumers in Amazon Prime without consent and sabotaging their attempts to cancel," *Press Release*, June 21, 2023. [Online]. Available: https://www.ftc.gov/news-events/news/press-releases/2023/06/ftc-takes-action-against-amazon-enrolling-consumers-amazon-prime-without-consent-sabotaging-their

[40] Federal Trade Commission, "FTC issues opinion finding TurboTax maker Intuit Inc. engaged in deceptive practices," *Press Release*, Jan. 22, 2024. [Online]. Available: https://www.ftc.gov/news-events/news/press-releases/2024/01/ftc-issues-opinion-finding-turbotax-maker-intuit-inc-engaged-deceptive-practices

[41] F. Westin and S. Chiasson, "Opt out of Privacy or "Go Home": Understanding Reluctant Privacy Behaviours through the FoMO-Centric Design Paradigm," in *Proc. NSPW*, San Carlos, Costa Rica, Sep. 2019, pp. 57–67.

[42] Federal Trade Commission, *Bringing Dark Patterns to Light: Staff Report*, Sept. 2022. [Online]. Available: https://www.ftc.gov/system/files/ftc_gov/pdf/P214800%20Dark%20Patterns%20Report%209.14.2022%20-%20FINAL.pdf

[43] E. Caragay and J. Zong, "Beyond Dark Patterns: A Concept-Based Framework for Ethical Software Design," in *Proc. CHI Conf. Hum. Factors Comput. Syst.*, Honolulu, HI, USA, May 2024, pp. 1–15.

[44] Federal Trade Commission, "Enforcement Policy Statement on Negative Option Marketing," *Fed. Regist.*, vol. 86, no. 205, pp. 59614—59619, Oct. 27, 2021.

[45] J. King and A. Stephan, "Regulating Privacy Dark Patterns in Practice— Drawing Inspiration from California Privacy Rights Act," *Geo. L. Tech. Rev.*, vol. 5, pp. 250–276, 2021.

[46] M. Goldberg et al., "Regulating privacy online: An economic evaluation of the GDPR," Am. Econ. J.: Econ. Policy, vol. 16, no. 1, pp. 325–358, 2024.

[47] G. Aridor, Y.-K. Che, and T. Salz, "The effect of privacy regulation on the data industry: Empirical evidence from the GDPR," *RAND J. Econ.*, vol. 54, no. 4, pp. 695–730, 2023.

[48] G. Johnson, T. Lin, J. C. Cooper, and L. Zhong, "COPPAcalypse? The YouTube settlement's impact on kids content," *SSRN*, May 24, 2024. [Online]. Available: https://papers.ssrn.com/sol3/papers.cfm?abstract_id=4430334

[49] R. Janssen et al., "GDPR and the lost generation of innovative apps," *Natl. Bur. Econ. Res.*, Working Paper Series, no. 30028, May 2022. [Online]. doi:10.3386/w30028. Available: https://www.nber.org/papers/w30028

[50] C. Chung, "TurboTax to refund $141 million in settlement over ads," *New York Times*, May 5, 2022, p. B3. [Online]. Available: https://www.propublica.org/article/inside-turbotax-20-year-fight-to-stop-americans-from-filing-their-taxes-for-free

[51] J. Elliott and P. Kiel,, "Inside TurboTax's 20-year fight to stop Americans from filing their taxes for free," *ProPublica*, Oct. 17, 2019. [Online]. Available: https://www.propublica.org/article/inside-turbotax-20-year-fight-to-stop-americans-from-filing-their-taxes-for-free